\documentclass[sigconf,screen]{acmart}

\usepackage{caption}
\usepackage{subcaption}
\usepackage{listings}
\usepackage{algorithm}
\usepackage{algpseudocode}

\usepackage{multirow}
\usepackage{bigdelim}
\usepackage{makecell}
\usepackage[utf8]{inputenc}
\usepackage[T1]{fontenc}
\usepackage{microtype}
\usepackage{setspace}

\newcommand{\code}[1]{\texttt{#1}}
\newcommand{\name}{LExecutor}

\newcommand{\CoverageFcts}{51.6\%} % open-source functions
\newcommand{\CoverageAsIsFcts}{4.1\%} % open-source functions
\newcommand{\CoverageImprovementFcts}{29\%} % open-source functions
\newcommand{\CoverageSO}{65.1\%} % Stack Overflow
\newcommand{\CoverageAsIsSO}{43.8\%} % Stack Overflow
\newcommand{\CoverageImprovementSO}{12.2\%} % Stack Overflow
\newcommand{\ModelAccuracyMin}{79.5\%}
\newcommand{\ModelAccuracyMax}{98.2\%}

\definecolor{halfgray}{gray}{0.35}
\definecolor{deepblue}{rgb}{0,0,0.5}
\definecolor{deepred}{rgb}{0.6,0,0}
\definecolor{deepgreen}{rgb}{0,0.5,0}
\definecolor{highlightorange}{rgb}{0.98, 0.92, 0.8}

\lstdefinelanguage{Python}{
	keywords={if, raise, print, and, or, is, None, not, in, elif, else, def, return <BUGGY_LINE>, <LINE>, <INITIAL_STATE>, <CONTEXT>, <DESIRED_STATE>},
	ndkeywords={},
	ndkeywordstyle=\color{darkgray}\bfseries,
	identifierstyle=\color{black},
	numberstyle=\color{deepred},
	sensitive=false,
	comment=[l]{\#},
	morestring=[b]',
	morestring=[b]"
}

\makeatletter
\newcommand\HL{%
	\gdef\lst@alloverstyle##1{%
		\fboxrule=0pt
		\fboxsep=0pt
		\colorbox{lightgray}{\strut##1}%
	}%
}
\newcommand\HLoff{%
	\xdef\lst@alloverstyle##1{##1}%
}

\ccsdesc[500]{Software and its engineering~Software testing and debugging}


\setcopyright{acmlicensed}
\acmPrice{15.00}
\acmDOI{10.1145/3611643.3616254}
\acmYear{2023}
\copyrightyear{2023}
\acmSubmissionID{fse23main-p119-p}
\acmISBN{979-8-4007-0327-0/23/12}
\acmConference[ESEC/FSE '23]{Proceedings of the 31st ACM Joint European Software Engineering Conference and Symposium on the Foundations of Software Engineering}{December 3--9, 2023}{San Francisco, CA, USA}
\acmBooktitle{Proceedings of the 31st ACM Joint European Software Engineering Conference and Symposium on the Foundations of Software Engineering (ESEC/FSE '23), December 3--9, 2023, San Francisco, CA, USA}
\received{2023-02-02}
\received[accepted]{2023-07-27}

\begin{document}

\title{\name{}: Learning-Guided Execution}

\author{Beatriz Souza}
\affiliation{%
  \institution{University of Stuttgart}
  \city{Stuttgart}
  \country{Germany}}
\email{beatrizbzsouza@gmail.com}

\author{Michael Pradel}
%\orcid{0000-0003-1623-498X}
\affiliation{%
  \institution{University of Stuttgart}
  \city{Stuttgart}
  \country{Germany}}
\email{michael@binaervarianz.de}

\begin{abstract}
  Executing code is essential for various program analysis tasks, e.g., to detect bugs that manifest through exceptions or to obtain execution traces for further dynamic analysis.
  However, executing an arbitrary piece of code is often difficult in practice, e.g., because of missing variable definitions, missing user inputs, and missing third-party dependencies.
  This paper presents \name{}, a learning-guided approach for executing arbitrary code snippets in an underconstrained way.
  The key idea is to let a neural model predict missing values that otherwise would cause the program to get stuck, and to inject these values into the execution.
  For example, \name{} injects likely values for otherwise undefined variables and likely return values of calls to otherwise missing functions.
  We evaluate the approach on Python code from popular open-source projects and on code snippets extracted from Stack Overflow.
 The neural model predicts realistic values with an accuracy between \ModelAccuracyMin{} and \ModelAccuracyMax{}, allowing \name{} to closely mimic real executions.
  As a result, the approach successfully executes significantly more code than any available technique, such as simply executing the code as-is.
  For example, executing the open-source code snippets as-is covers only \CoverageAsIsFcts{} of all lines, because the code crashes early on, whereas \name{} achieves a coverage of \CoverageFcts{}.
\end{abstract}

\maketitle

\section{Introduction}

The ability to execute code enables various dynamic program analysis applications.
For example, running a program may expose bugs that trigger obviously wrong behavior, such as runtime exceptions.
Moreover, dynamic analysis has been shown its usefulness for various tasks, e.g., mining API protocols~\cite{Gabel2008,Yang2006,ase2009,DBLP:journals/tse/RobillardBKMR13}, taint analysis~\cite{Clause2007,DBLP:conf/sp/SchwartzAB10}, and type inference~\cite{An2011,DBLP:conf/sac/RenTSF13}.
Generally, dynamic analysis is well known to complement static analysis~\cite{Ernst2003,Smaragdakis2007}, providing a valuable technique for understanding and improving software.

Unfortunately, executing arbitrary code is challenging, both at the small and the large scale.
Small-scale code snippets, e.g., posted on Stack Overflow,
often are missing contextual information, such as imports and definitions of variables and functions~\cite{DBLP:journals/corr/abs-1907-04908}.
Large-scale code bases often are non-trivial to set up and run, e.g., due to missing third-party dependencies, complex build procedures, and the lack of inputs that reach deeply into the project.
Overall, the difficulties of executing code limit the potential of dynamic analysis.

\begin{figure}
  \includegraphics[width=\linewidth]{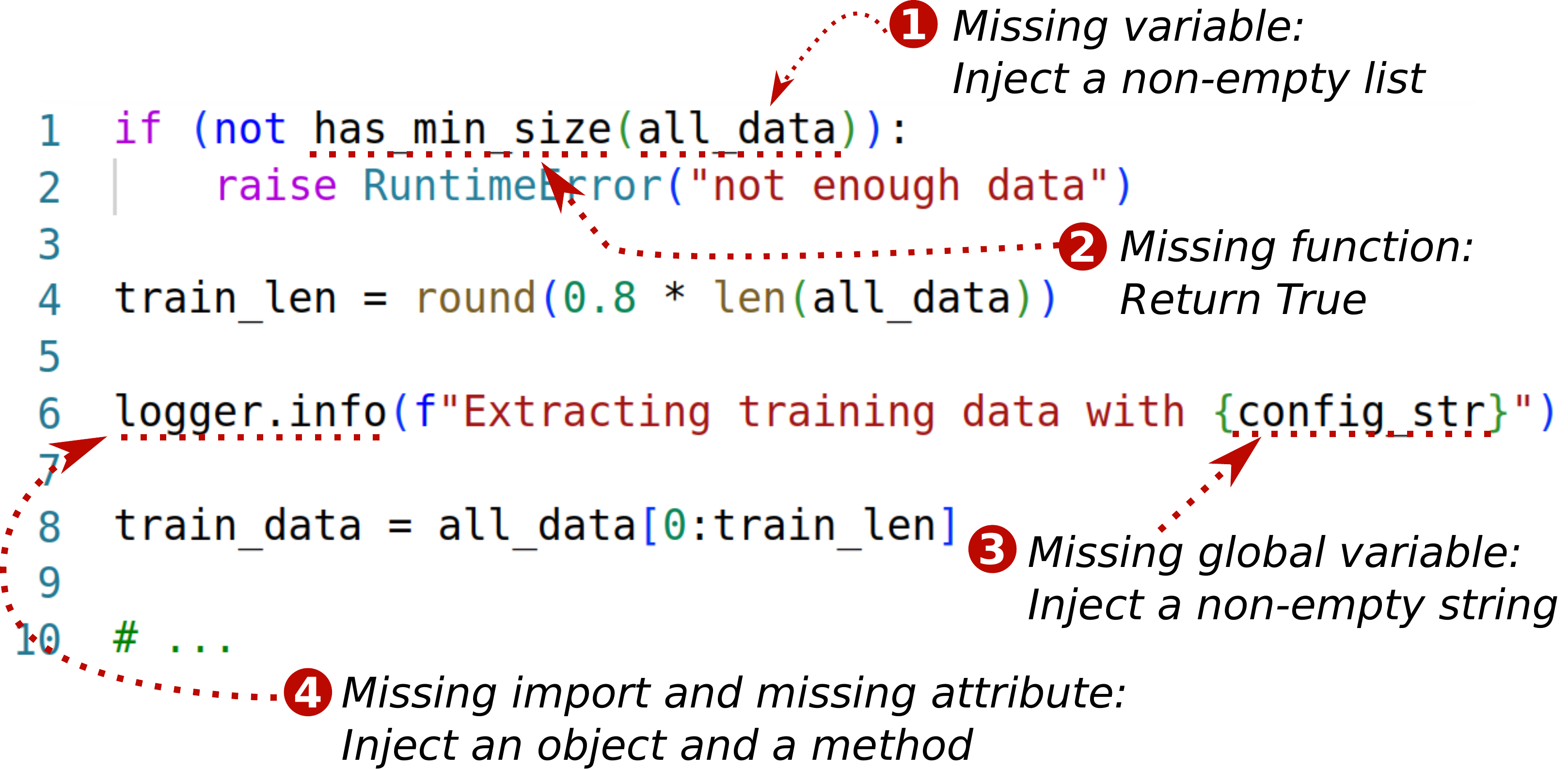}
  \caption{Python code to execute and how \name{} guides the execution.}
  \label{fig:example}
  \vspace{-1.7em}
\end{figure}

For example, consider the Python code snippet in Figure~\ref{fig:example}, which could be the body of a function extracted from a complex code base or code posted on Stack Overflow.
Executing this code is difficult because of various missing pieces of contextual information.
At first, the code tries to read variable \code{all\_data}, which does not exist, and hence, will cause the code to crash.
Even we had a definition for this variable, then other obstacles would prevent the code from executing, such as the missing function \code{has\_min\_size}, the missing (global) variable \code{config\_str}, and the missing imported method \code{logger.info}.
A skilled human can intuitively fill in the missing information by predicting it to be likely that, e.g., \code{all\_data} is a list, \code{has\_min\_size} will return a boolean, \code{config\_str} is a string, and that \code{logger.info} is a method call likely to succeed.
Given this information, a human could mentally emulate the execution and hence reason about the runtime behavior of the code.
% Of course, manually reasoning about code in this manner does not scale well to analyzing larger amounts of code.
The key question we ask in this work is: \emph{Can we automate the prediction of likely values and use them to execute otherwise non-executable code?}

This paper presents \name{}, which offers a learning-guided approach for executing arbitrary code snippets in an underconstrained way.
The key idea is to let a neural model predict suitable values whenever the program usually would be stuck.
The approach is enabled by three components:
\begin{itemize}
  \item A specifically designed and trained neural learning model that predicts runtime values based on the code context in which a value is used.
  We realize this component by fine-tuning a model with hundreds of thousands of examples gathered from regular executions of real-world programs.
  \item An execution environment that prevents crashes due to missing values and instead fills in model-predicted values.
  \item An AST-based instrumentation technique that turns arbitrary Python code into ``lexecutable'' code.
  % The instrumentation is realized as an AST-based source-to-source transformation.
\end{itemize}

For the example in Figure~\ref{fig:example}, \name{} prevents the program from crashing and instead predicts suitable values for all undefined variables and functions.
The figure shows what values the approach predicts and then injects into the execution.
For example, when the code tries to access the otherwise missing value \code{has\_min\_size}, then \name{} predicts it to be a function that returns the boolean value \code{True}.
Given the injected values, the code successfully executes without prematurely terminating.
There is no guarantee that the values that \name{} injects exactly match those that would occur in a ``real'' execution of the given code.
However, we find that \name{} is able to predict realistic values in \emph{most} situations where a regular execution would simply get stuck, enabling the execution of otherwise non-executable code.

Our work has overlapping goals with four prior streams of research:
(1) Test generators~\cite{Pacheco2007,Fraser2011a}, as they also provide missing input values.
(2) Concolic execution~\cite{Godefroid2005,Sen2005,Cadar2008}, which abstracts missing inputs into symbolic variables and constraints.
(3) Micro-execution~\cite{Godefroid2014} and underconstrained symbolic execution~\cite{Ramos2015}, which execute arbitrary code by injecting random values into memory on demand.
In contrast to (1) and (2), \name{} injects otherwise missing values at arbitrary points within an execution, whereas the prior work provides values only at well-defined interfaces, e.g., command-line arguments or method arguments.
In contrast to all the above work, \name{} uses a neural model to predict realistic values, whereas (1) and (3) inject random values, and (2) overapproximates possible values via symbolic reasoning.
(4) Neural models that predict various properties of code~\cite{Raychev2015,NeuralSoftwareAnalysis}.
In contrast to (4), \name{} predicts runtime values during an execution, whereas prior work focuses on predicting static properties of code.

We evaluate our work by applying it to two sets of code snippets:
functions extracted from popular open-source projects and code snippets extracted from Stack Overflow posts.
As baselines, the evaluation compares \name{} to regular execution, injecting random values, and a state-of-the-art, function-level test generator for Python~\cite{DBLP:conf/ssbse/LukasczykKF20}.
We show that the neural model at the core of \name{} predicts realistic values with an accuracy of \ModelAccuracyMin{} to \ModelAccuracyMax{} (depending in the configuration).
Moreover, our results show that the approach enables the execution of \CoverageFcts{} and \CoverageSO{} of all lines in the open-source code and Stack Overflow snippets, respectively, which improves over the best baseline by \CoverageImprovementFcts{} and \CoverageImprovementSO{}.
Finally, we apply \name{} to detect semantics-changing commits by comparing the execution behavior of a function before and after a commit.

We envision \name{} to enable various dynamic analysis applications.
For example, ``lexecuting'' code could help detect bugs that manifest through obvious signs of misbehavior, such as runtime exceptions or assertion violations.
Likewise, our approach could be used to validate code generated by code synthesis techniques~\cite{Ferdowsifard2021} or generative language models~\cite{Chen2021,Xu2022}.
Other potential applications include to check if and how a code change to modifies the observable behavior~\cite{Ramos2015}, and classical dynamic analyses, such as detecting security vulnerabilities via taint analysis.

In summary, this paper contributes the following:
\begin{itemize}
  \item A novel way of executing arbitrary code in an underconstrained and learning-guided way.
  \item A neural model designed and trained to predict realistic values to be injected into an execution. The task addressed by the model fundamentally differs from prior work on predicting various properties of code~\cite{Raychev2015,NeuralSoftwareAnalysis}, because we here predict runtime values.
  \item An instrumentation-based runtime environment that prevents programs from crashing due to missing values and instead injects values provided by the model.
  \item Empirical evidence that the approach effectively injects realistic values and outperforms several baselines at covering and successfully executing code.
\end{itemize}

\section{Overview}

\begin{figure}
  \includegraphics[width=.8\linewidth]{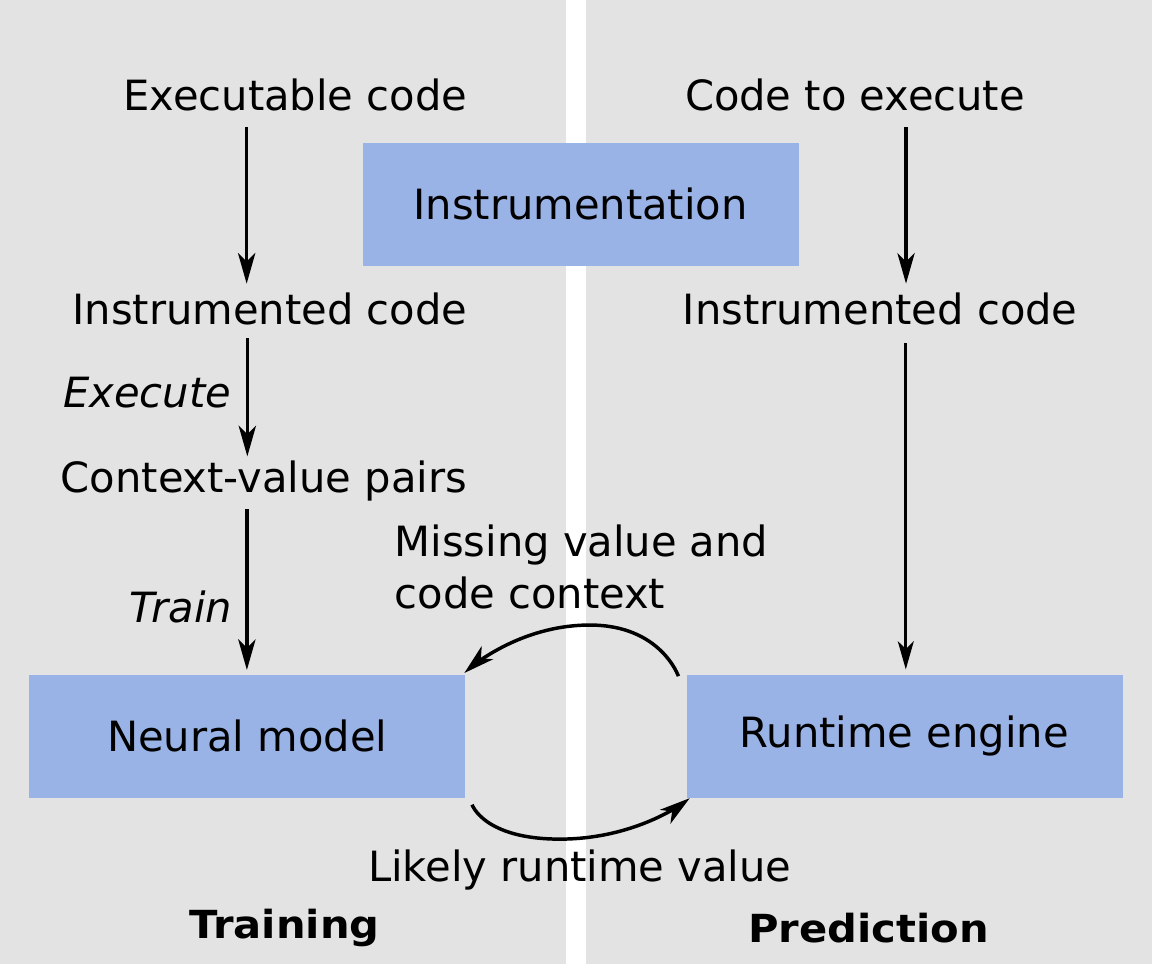}
  \caption{Overview of learning-guided execution.}
  \label{fig:overview}
\end{figure}

Figure~\ref{fig:overview} gives an overview of our approach.
\name{} has a \emph{training phase} and a \emph{prediction phase}.
The training phase yields a \emph{neural model} that, given the code context in which a value is used during an execution, predicts a suitable value.
To this end, we train a model based on pairs of code context and a value used in this context, which we gather by executing a corpus of code.
Once the neural model is trained, \name{} is ready for the prediction phase.
The input to the approach is a possibly incomplete piece of code to execute.
To execute this code despite possibly missing information, a \emph{runtime engine} intercepts any use of a value, such as a read of a variable.
If the value is not defined, i.e., the program would usually terminate prematurely, the runtime engine queries the neural model and then injects the value proposed by the model into the execution.

Both training and prediction are enabled by source-to-source \emph{instrumentation}, which takes regular Python code and adds new instructions to it.
The added instructions wrap all uses of values into special \emph{value loaders} that implement the functionality of \name{} while otherwise preserving the original semantics.

For the example in Figure~\ref{fig:example}, an execution starts at line~1 by using two values, first \code{all\_data} and then \code{has\_min\_size}.
\name{} wraps both into value loaders that allow the runtime engine to observe that these values are missing, and to hence inject values that would likely occur in a regular execution.
As shown in the figure, the approach injects a non-empty list for \code{all\_data} and a function returning \code{True} for \code{has\_min\_size}.
As a result, the code continues to execute, passes line~4 without any need for guidance by \name{}, and then reaches line~6.
Here, several values are missing, which the runtime engine provides successfully.
For the missing method \code{logger.info}, the approach first injects a dummy object for \code{logger}, then resolves the missing attribute \code{info} into a function, and finally injects the return value \code{None} when calling the function.
By guiding the execution in this way, \name{} successfully executes all code shown in the figure.

\section{Approach}

The following presents in detail the three main components of \name{}: the code instrumentation, the neural model, and the runtime engine.

\subsection{Code Instrumentation}
\label{sec:instr}

The goal of the instrumentation is to observe runtime values to be used during the training phase and to inject otherwise missing values during the prediction phase.
We perform a series of AST-based code transformations via a top-down, left-to-right pass over all nodes in the given AST.
The pass transforms three kinds of AST nodes and the corresponding AST subtrees under them:
\begin{itemize}
  \item \emph{Variable reads}. The instrumentation transforms all nodes that refer to the name of a variable that gets read. In contrast, we do not instrument variable writes. For example, in \code{y = x + 1}
  the reference to \code{x} gets instrumented, as it might read an undefined value, which \name{} tries to prevent.
  In contrast, the assignment to \code{y} is not instrumented and hence executed as in the original code.
  \item \emph{Attribute reads}. The instrumentation transforms all nodes that refer to an attribute that gets read. Again, we do not instrument writes of attributes. For example, in \code{y.foo = x.bar} both the read of the variable \code{x} (as described above) and the read of the attribute \code{bar} get instrumented, because they might access otherwise undefined values.
  \item \emph{Calls of functions and methods}. The instrumentation transforms all calls of functions and methods (we say ``functions'' to mean both). For example, in
    \code{y = foo()}
  the call of \code{foo} gets instrumented so \name{} can inject a return value if the function is undefined.
 \end{itemize}

Each of the three kinds of instrumented nodes is replaced by a call to a special \emph{loader function}.
For variable reads, the instrumented code calls a loader function \code{\_n\_} and passes two pieces of information to it: the name of the variable and a newly created lambda function that tries to read and then return the value of the variable.
The lambda function enables \name{} to try to read the variable in a controlled manner and to react accordingly depending on whether the value is available.
For attribute reads, the instrumented code passes two pieces of information to a loader function \code{\_a\_}: the value of the base object (which when accessing the attribute has already been evaluated) and the name of the attribute.
The loader function will then try to read the attribute of the base object and react accordingly depending on whether the attribute is available.
Finally, for function calls, the instrumented code passes the callee (i.e., the called function) to a loader function \code{\_c\_}.
The loader function will then invoke the callee, while handling cases where the function is undefined.
In addition to the parameters mentioned above, each of the loader functions also receives a unique identifier of the instrumented source code location, which \name{} later uses to access the code context of the location.

\begin{figure}
  \emph{Original code:}\\
\begin{lstlisting}
x = foo()
y = x.bar + z
\end{lstlisting}

  \emph{Instrumented code:}\\
\begin{lstlisting}
x = _c_(536, _n_(535, "foo", lambda: foo))
y = _a_(538, _n_(537, "x", lambda: x), "bar") \
        + _n_(539, "z", lambda: z)
\end{lstlisting}

  \caption{Example of code instrumentation.}
  \label{fig:instr ex}
\end{figure}

Figure~\ref{fig:instr ex} illustrates the instrumentation with an example, where the original and the instrumented code are shown at the top and bottom, respectively.
The call of \code{foo()} in the first line consists of two separately instrumented instructions: reading the variable \code{foo}, which gets wrapped into \code{\_n\_}, and calling the function stored in this variable, which gets wrapped into \code{\_c\_}.
The numbers passed as the first argument to the loader functions are the unique identifiers of code locations.
The second line of the original code contains three instructions to instrument: the two variable reads of \code{x} and \code{z}, and the attribute access wrapped into \code{\_a\_}.
Note that the calls to the loader functions are nested into each other.
For example, in the first instrumented line, the value returned by \code{\_n\_} as the result of reading variable \code{foo} is passed as the callee to \code{\_c\_}.

In addition to the instrumentation described above, the approach adds to each instrumented file imports of the loader functions.
With these imports, the instrumented file serves as a drop-in replacement of the original file.
Apart from changing the uses of values, the instrumentation does not modify the semantics of the code.

\subsection{Neural Model}

We next describe the neural model, which predicts likely runtime values to use in a given code context.
% The rationale for choosing a neural model is the effectiveness of deep learning models at code-related prediction tasks, as evidenced by a stream of recent work on neural software analysis~\cite{NeuralSoftwareAnalysis}.

\subsubsection{Gathering Training Data}

To gather data for training the model we execute and dynamically analyze tests suites of popular Python projects.
The training executions are not guided by \name{}, but regular executions with all inputs and dependencies given.
% As a scalable way of executing large amounts of real-world code, we use the tests suites of popular Python projects.
%
Via dynamic analysis the approach gathers a trace of value-use events:

\begin{definition}\label{def:value-use event}
  A \emph{value-use event} is a tuple $(n, v, k, l)$, where
  \begin{itemize}
    \item $n$ is the name of the used reference, i.e., a variable name, an attribute name, or the name of a function;
    \item $v$ is an abstracted version of the value that gets used;
    \item $k$ indicates the kind of used value, which is either \emph{variable}, \emph{attribute}, or \emph{return value}; and
    \item $l$ is a unique identifier of the code location where the value gets used.
  \end{itemize} 
\end{definition}

To gather a trace of such events, \name{} instruments all executed code (Section~\ref{sec:instr}).
The loader functions invoked by the instrumented code load the requested value, as the original code, and in addition, add a value-use event to the trace.
At the end of the execution, \name{} stores the recorded trace into a file.

\subsubsection{Representing Values for Learning}
\label{sec:value representation}

Given traces of value-use events, we need to represent these data in a format suitable for deep learning.
An important part of this step is to abstract the concrete runtime values observed during an execution into a finite number of classes, which will be the prediction targets of the neural model.
The abstraction has two conflicting goals.
On the one hand, we aim for a fine-grained, precise prediction of what value to inject into an execution.
For example, if the code uses a variable \code{age}, predicting the exact integer to load would enable \name{} to produce the most realistic possible execution.
On the other hand, we aim for a highly accurate model, which becomes easier if there are only few, coarse-grained classes of values to choose from.
For the above example, such a coarse-grained prediction could simply say that the value should be an integer, without any more information about the specific value.

\begin{table}
  \caption{Fine-grained abstraction and concretization of values.}
  \label{tab:value abstraction}
  \setlength{\tabcolsep}{28pt}
  \small
  \begin{tabular}{@{}ll@{}}
    \toprule
    Abstract class of values & Concretization (Python) \\
    \midrule
    
    \multicolumn{2}{@{}l@{}}{\emph{Common primitive values:}} \\
    \hspace{.8em} None & \code{None} \\
    \hspace{.8em} True & \code{True} \\
    \hspace{.8em} False & \code{False} \\

    \multicolumn{2}{@{}l@{}}{\emph{Built-in numeric types:}} \\
    \hspace{.8em} Negative integer & \code{-1} \\
    \hspace{.8em} Zero integer & \code{0} \\
    \hspace{.8em} Positive integer & \code{1} \\
    \hspace{.8em} Negative float & \code{-1.0} \\
    \hspace{.8em} Zero float & \code{0.0} \\
    \hspace{.8em} Positive float & \code{1.0} \\

    \multicolumn{2}{@{}l@{}}{\emph{Strings:}} \\
    \hspace{.8em} Empty string & \code{""} \\
    \hspace{.8em} Non-empty string & \code{"a"}\\
    
    \multicolumn{2}{@{}l@{}}{\emph{Built-in sequence types:}} \\
    \hspace{.8em} Empty list & \code{[]} \\
    \hspace{.8em} Non-empty list & \code{[Dummy()]} \\
    \hspace{.8em} Empty tuple & \code{()} \\
    \hspace{.8em} Non-empty tuple & \code{(Dummy())} \\

    \multicolumn{2}{@{}l@{}}{\emph{Built-in set and dict types:}} \\
    \hspace{.8em} Empty set & \code{set()} \\
    \hspace{.8em} Non-empty set & \code{set(Dummy())} \\
    \hspace{.8em} Empty dictionary & \code{\{\}} \\
    \hspace{.8em} Non-empty dictionary & \code{\{"a": Dummy()\}} \\

    \multicolumn{2}{@{}l@{}}{\emph{Functions and objects:}} \\
    \hspace{.8em} Callable & \code{Dummy} \\
    \hspace{.8em} Resource & \code{DummyResource()} \\
    \hspace{.8em} Object & \code{Dummy()} \\

    \bottomrule
  \end{tabular}
\end{table}

\name{} strikes a balance between these two conflicting goals by using a fixed-size set of abstracted values.
By default, this set comprises 23 classes as shown in Table~\ref{tab:value abstraction}.
The left column gives the classes of values that the neural model distinguishes.
For some very common primitive values, namely \code{None}, \code{True}, and \code{False}, the approaches preserves the concrete values, enabling \name{} to predict them exactly.
For numeric types, such as \code{int} and \code{float}, we abstract the concrete values into three classes per type, which represent negative, zero, and positive values, respectively.
To represent strings and common data structure types, such as lists, the approach distinguishes between empty and non-empty values, e.g., an empty list as opposed to a non-empty list.
All callable values, such as functions, are represented as a class \emph{callable}.
The \emph{resource} class represents values that can be opened and closed, such as file pointers, which are often used with the \code{with} keyword in Python.
Finally, all remaining non-primitive types are abstracted as \emph{object}.
We refer to the value abstraction described above as \emph{fine-grained}.

\begin{table}
  \caption{Coarse-grained abstraction and two modes for concretizing values.}
  \label{tab:coarse-grained value abstraction}
  \setlength{\tabcolsep}{11pt}
  \small
  \begin{tabular}{@{}lll@{}}
    \toprule
    Abstract class & \multicolumn{2}{c}{Concretization (Python)} \\
    \cmidrule{2-3}
    of values & Deterministic & Randomized \\

    \midrule
    
    \multicolumn{3}{@{}l@{}}{\emph{Common primitive values:}} \\
    \hspace{.8em} None & \code{None} \\
    \hspace{.8em} Boolean & \code{True} & \code{True}, \code{False} \\

    \multicolumn{3}{@{}l@{}}{\emph{Built-in numeric types:}} \\
    \hspace{.8em} Integer & \code{1} & \code{-1}, \code{0}, \code{1} \\
    \hspace{.8em} Float & \code{1.0} & \code{-1.0}, \code{0.0}, \code{1.0} \\

    \multicolumn{3}{@{}l@{}}{\emph{Strings:}} \\
    \hspace{.8em} String & \code{"a"} & \code{""}, \code{"a"}\\
    
    \multicolumn{3}{@{}l@{}}{\emph{Built-in sequence types:}} \\
    \hspace{.8em} List & \code{[Dummy()]} & \code{[]}, \code{[Dummy()]} \\
    \hspace{.8em} Tuple & \code{(Dummy())} & \code{()}, \code{(Dummy())} \\

    \multicolumn{3}{@{}l@{}}{\emph{Built-in set and dict types:}} \\
    \hspace{.8em} Set & \code{set(Dummy())} & \code{set()}, \code{set(Dummy())} \\
    \hspace{.8em} Dictionary & \code{\{"a": Dummy()\}} & \code{\{\}}, \code{\{"a": Dummy()\}} \\

    \multicolumn{3}{@{}l@{}}{\emph{Functions and objects:}} \\
    \hspace{.8em} Callable & \code{Dummy} \\
    \hspace{.8em} Resource & \code{DummyResource()} \\
    \hspace{.8em} Object & \code{Dummy()} \\

    \bottomrule
  \end{tabular}
\end{table}

To better understand the impact of the number of classes that values are abstracted into, \name{} also supports a more \emph{coarse-grained} value abstraction with only 12 classes, as shown in Table~\ref{tab:coarse-grained value abstraction}.
Instead of distinguishing between different values of a type, the coarse-grained value abstraction maps all values of a type into a single class.
For example, the abstraction does not distinguish between negative, zero, and positive integers, but simply represents all of them as integers.
Unless otherwise mentioned, all results are based on the fine-grained variant of the approach.

\subsubsection{Representing Code Context for Learning}
\label{sec:context representation}

The goal of the neural model is to predict one of the abstract value classes for each missing value.
To this end, we provide the following contextual information as an input to model:

\begin{definition}
  \label{def:model input}
  The \emph{input to the model} is a sequence of tokens
  $$n~\langle{}sep\rangle{}~k~\langle{}sep\rangle{}~c_{pre}~\langle{}mask\rangle{}~c_{post}$$ where
  \begin{itemize}
    \item $n$ is the name used to refer the to-be-predicted value;
    \item $\langle{}sep\rangle{}$ is a special separator token
    \item $k$ is the kind of value to predict, i.e., \emph{variable}, \emph{attribute}, or \emph{return value}.
    \item $c_{pre}$ are the code tokens just before the reference to the to-be-predicted value;
    \item $\langle{}mask\rangle{}$ is a special masking token; and
    \item $c_{post}$ are the code tokens just after the reference to the to-be-predicted value.
  \end{itemize}
\end{definition}

The prediction task is related to the well-known unmasking task, which is commonly used to train language models~\cite{Devlin2018,Feng2020}.
In contrast to unmasking, we do not ask the model to predict a masked token, which would be trivial, as it is given in $n$ as part of the input.
Instead, we ask the model to predict what value the masked token most likely evaluates to.

\paragraph{Example}
Suppose the code in Figure~\ref{fig:example} gets executed with all required contextual information, e.g., as part of a larger project, and that we gather training data from its execution.
The approach keeps track of all used values, such as the read of \code{all\_data} at line~1.
Following Definition~\ref{def:value-use event}, the approach records a value-use event $(n, v, k, l)$ with $n=$``all\_data'', $v=$``non-empty list'', $v=$``variable'', and $l$ being a unique identifier of the code location.
The corresponding input to the model is a tokenized version of the following:
\begin{gather*}
  \code{all\_data}~
  \langle{}sep\rangle{}~
  \emph{variable}~
  \code{if (not has\_min\_size(}~ 
  \langle{}mask\rangle{}~ \\
  \code{)): raise RuntimeError( ...}  
\end{gather*}
The prediction target for this example is ``non-empty list''.

\subsubsection{Training and Prediction}

The data preparation steps described above yield pairs of input sequences and abstracted values.
Before training the model, we deduplicate these pairs, which greatly reduces the amount of training data because programs often repeatedly load the same value at the same location, e.g., in a loop.
We then train a single model for all three kinds of values to predict, allowing the model to generalize across values stores in variables and attributes, as well as return values of function calls.

For the neural model, our approach can build on any model that accepts a sequence of tokens as the input and then either acts as a classifier or predicts tokens from a given vocabulary.
Our current implementation integrates two pre-trained models, CodeT5~\cite{DBLP:conf/emnlp/0034WJH21} and CodeBERT~\cite{Feng2020}, which we fine-tune for our prediction task.
Building upon a pre-trained model enables our approach to benefit from this model's abilities in ``understanding'' both code and natural language.
CodeT5 uses the T5 architecture~\cite{Raffel2020}, a transformer-based neural network architecture that maps a sequence of input tokens to a sequence of output tokens.
CodeBERT follows the BERT~\cite{Devlin2018} architecture and is pre-trained to unmask missing tokens and predict whether a token has been replaced.
For both models, \name{} tokenizes the input and output using the tokenizer that comes with the respective model.
To fit the input sequence into the fixed input size of the models (512 tokens), we truncate and pad the code tokens in $c_{pre}$ and $c_{post}$ (Definition~\ref{def:model input}).
For training, we use the Adam optimizer with weight decay fix~\cite{DBLP:conf/iclr/LoshchilovH19}.
We leave all hyperparameters at the defaults, except for the batch size, which we reduce to 50 (CodeT5) and 13 (CodeBERT) so it fits into our available GPU memory.

\subsection{Runtime Engine}
\label{sec:runtime engine}

Once the model is trained, the prediction phase of \name{} executes possibly incomplete code via learning-guided execution.
The core of this step is the runtime engine, which intercepts all loaded values and injects otherwise missing values on demand.
To this end, the runtime engine implements the loader functions added to the original code during the instrumentation (Section~\ref{sec:instr}).
The basic idea is to load and return the original value whenever possible, and to fall back on querying the model otherwise.

\begin{algorithm}[t]
  \caption{Value loading and injection in the runtime engine.}
  \label{alg:runtime}
  \begin{algorithmic}[1]
    \Require Kind $k$ of value, model input $i$
    \Ensure Concrete value $v$

    \If{$k =$ ``name''}\label{line:name start}
      \State $v \leftarrow$ load value or catch \code{NameError}\label{line:load and catch}
      \If{no exception while loading the value}:
        \State \textbf{return} $v$
      \EndIf
    \ElsIf{$k =$ ``attribute''}
      \State $v \leftarrow$ load value or catch \code{AttributeError}
      \If{no exception while loading the value}:
        \State \textbf{return} $v$
      \EndIf\label{line:attribute end}
    \ElsIf{$k =$ ``return value''}\label{line:call start}
      \If{callee $f$ is a regular function}
        \State \textbf{return} $v \leftarrow f()$
      \EndIf
    \EndIf\label{line:call end}
    \State $v_{abstr} \leftarrow \mathit{model}(i)$\label{line:call model}
    \State $v \leftarrow \mathit{concretize}(v_{abstr})$\label{line:concretize}
    \State \textbf{return} $v$
  \end{algorithmic}
\end{algorithm}

Algorithm~\ref{alg:runtime} presents how the engine loads and, if needed, injects values.
For uses of variables and attributes (lines~\ref{line:name start} to~\ref{line:attribute end}), the runtime engine tries to read the value while catching any exception thrown when a variable or attribute is undefined.
If reading the value succeeds, then this value is returned and the regular execution of the code continues.
Otherwise, the engine will predict and inject a value, as described below.
For calls of functions (lines~\ref{line:call start} to~\ref{line:call end}), the algorithm distinguishes two cases.
If the callee is a regular function, i.e., a function that was successfully resolved without any help by \name{}, then the runtime engine simply calls this function and returns its return value.
If, instead, the function is a dummy value that was injected by \name{} because the code tried to read a function value that does not exist, then the engine predicts and injects a suitable return value.

Whenever Algorithm~\ref{alg:runtime} reaches line~\ref{line:call model}, it has failed to load a regular value.
In a regular execution, the code would terminate prematurely in this case.
Instead, the algorithm queries the neural model for a suitable value to inject, and the model returns one of the abstract classes in Tables~\ref{tab:value abstraction} and~\ref{tab:coarse-grained value abstraction}.
To enable the code to continue its execution, the runtime engine concretizes the abstract class into a runtime value.
In our default configuration, i.e., using the fine-grained value abstraction, the concrete values are those shown in the right column of Table~\ref{tab:value abstraction}.
They are simple default values, such as \code{-1} for negative integers or \code{"a"} for non-empty strings.
Whenever the engine injects an object, it creates a new instance of an empty dummy class \code{Dummy}.
For injecting callables, i.e., in situations where the code tries to access a non-existing function, the engine returns the constructor of the \code{Dummy} class.
The reason is that constructors are the most common type of callable in our dataset.

When using the coarse-grained value abstraction, \name{} supports two modes for deciding what concrete values to inject.
In \emph{deterministic} mode, the approach always injects the value shown in the middle column of Table~\ref{tab:coarse-grained value abstraction}.
In contrast, in \emph{randomized} mode, the approach randomly picks from the values shown in the right column of the same table.
For example, when the model predicts a string, then the deterministic mode always provides the value \code{"a"}, whereas randomized mode returns either \code{""} or \code{"a"}.
To avoid executions that normally would be be impossible, the randomized mode ensures that repeatedly using the same variable returns a consistent value.
Future work could further improve the value concretization, e.g., by trying to predict meaningful string values or by predicting the types of values a non-empty list should contain.

Importantly, the runtime engine does not prevent all runtime errors that might occur during an execution.
Instead, \name{} prevents only those exceptions that would be caused by trying to use a non-existing value.
That is, code executed with \name{} may still fail, e.g., due to some logical bug in the 
code.

\paragraph{Example}
Suppose that we are trying to execute the code in Figure~\ref{fig:example} during \name{}'s prediction mode.
When reaching the use of \code{all\_data} at line~1, Algorithm~\ref{alg:runtime} attempts to load the value of the variable, which leads to a \code{NameError} because the variable is undefined.
The runtime engine catches this error and instead queries the neural model for a suitable value to inject.
The model predicts ``non-empty list'', which Algorithm~\ref{alg:runtime} concretizes, as shown in Table~\ref{tab:value abstraction}, into a list that contains a dummy object.
This injected value along with others injected later during the execution enable the code to execute despite some missing values.

\section{Implementation}

\name{} is implemented as a fully automated tool that executes arbitrary Python code snippets.
To implement the code instrumentation we build upon the LibCST library,\footnote{\url{https://github.com/Instagram/LibCST}} which offers a parser and utilities for transforming the AST.
The neural models are based on CodeT5 made available by Salesforce\footnote{\url{https://github.com/salesforce/CodeT5}} and CodeBERT made available by Microsoft\footnote{\url{https://github.com/microsoft/CodeBERT}}, both accessed via the Hugging Face API.
Due to hardware constraints, we use the ``small'' CodeT5 model and the ``base-mlm'' CodeBERT model, and fine-tune them for ten epochs each.
To speed up the prediction of values, our implementation loads the trained model once and accesses it via REST queries to an HTTP server.
The experiments are conducted on three servers, each with an NVIDIA Tesla GPU (P100, T100, and T4, respectively) using 16GB of GPU memory per server.
Deploying \name{} does not require a powerful server, except for running the neural model.

\section{Evaluation}

We address the following research questions:
\begin{itemize}
\item RQ1: How accurate is the neural model at predicting realistic runtime values?
% To answer this question, we compare the values predicted by the model against values observed during regular executions of real-world Python projects.
\item RQ2: How much code does an execution guided by \name{} cover, and how does the coverage compare to alternative ways of executing the code?
% To answer this question, we apply \name{} to thousands of code snippets and compare the number of successfully executed lines of code against baselines, such as simply trying to execute the code as-is.
\item RQ3: How efficient is ``lexecuting'' code?
% We measure how long it takes to execute an average line of code, including the time to fill in missing values.
\item RQ4: As an application of the approach, can we use \name{} to identify semantics-changing commits?
\end{itemize}

\subsection{Experimental Setup}

\subsubsection{Datasets}
\label{datasets}

\begin{table}
  \caption{Projects used for gathering training data.}
  \label{tab:trace projects}
  \setlength{\tabcolsep}{7pt}
  \small
  \begin{tabular}{@{}llr@{}}
    \toprule
    Project & Description & Unique \\
    && value-use \\
    && events \\
    \midrule
    Ansible & Automation of software infrastructure & 43,090 \\
    Django & Web framework & 121,567 \\
    Keras & Deep learning library & 30,709 \\
    Requests & HTTP library & 5,273 \\
    Rich & Text formatting in the terminal & 25,370 \\
    \midrule
    Total & & 226,009 \\
    \bottomrule
  \end{tabular}
\end{table}

\paragraph{Value-use events}
To gather a corpus of value-use events for training the neural model, we execute the test suites of real-world Python projects.
We select projects that are popular (based on the number of GitHub stars), cover different application domains, and provide a test suite that is easy to execute yet covers a significant amount of the project's code.
Table~\ref{tab:trace projects} shows the selected projects and the number of unique value-use events gathered from them.
In total, the dataset consists of 226k entries, which we shuffle and then split into 95\% for training and 5\% to answer RQ1.

\begin{table}
  \caption{Projects used for gathering functions.}
  \label{tab:random_functions_projects}
  \setlength{\tabcolsep}{13pt}
  \small
  \begin{tabular}{@{}llrr@{}}
    \toprule
    Project & Description & Functions & LoC \\
    \midrule
    Black & Code formatting & 200 & 2,961 \\
    Flask & Web applications & 200 & 1,354 \\ 
    Pandas & Data analysis & 200 & 2,015 \\
    Scrapy & Web scraping & 200 & 1,198 \\
    TensorFlow & Deep learning & 200 & 2,125 \\
    \midrule
    Total && 1,000 & 9,653 \\
    \bottomrule
  \end{tabular}
\end{table}

\paragraph{Open-source functions}
One usage scenario of \name{} is to execute code that is part of a large project, without having to set up the project and its dependencies, and without having to provide inputs that reach the targeted code.
To evaluate \name{}'s effectiveness in this scenario, we gather a dataset of functions extracted from open-source Python projects on GitHub.
We search for projects implemented in Python and sort them by the number of their stars.
Then, from the top of the list we choose projects that cover different application
domains and that are not used for constructing the dataset of value-use events.
Then, for each of the projects, listed in Table~\ref{tab:random_functions_projects}, we randomly select 200 functions and extract the entire function body into a separate file.
The task of \name{} is to execute the extracted code without access to any contextual information, such as other code in the same project, imports, or third-party libraries.
In total, the dataset is composed of 1,000 randomly selected functions, that amount to 9,653 non-empty, non-comment lines of code.

\paragraph{Stack Overflow snippets}
Another usage scenario is to execute code snippets that are inherently incomplete, such as code posted in web forums.
To evaluate \name{} in this scenario, we gather a dataset of code snippets from Stack Overflow.
Specifically, we search for questions tagged with Python and sort them by the number of votes. Then, from each of the top 1,000 questions, we randomly select an answers and extract the code given in this answer.
After discarding code snippets with invalid syntax, the final dataset consists of 462 code snippets.
The total number of non-empty, non-comment lines of code is 3,580 lines.

\subsubsection{Baselines}
\label{baselines}

\paragraph{As-is}
This baseline executes a code snippet with the standard Python interpreter, i.e., without making any value predictions, and instead letting the code crash whenever it tries to access a missing value.

\paragraph{Naive value predictor}
This and the following two baselines are simplified variants of our approach, which use the runtime engine to intercept exceptions caused by undefined references and then inject values instead. The naive value predictor always predicts \emph{object}, and hence always injects a freshly created dummy object.
  
\paragraph{Random value predictor}
This baseline randomly samples (with uniform distribution) a value from the 23 classes of abstracted values available in Table~\ref{tab:value abstraction}.
  % Conceptually, this baseline is similar to micro-execution~\cite{Godefroid2014}, which injects random bytes into memory, but implemented for Python instead of x86 binaries.
  
\paragraph{Frequency-based value predictor}
This baseline associates names of variables, attributes, and functions in the training dataset with their distribution of abstract values.
  When asked for a value for a particular name, it samples from the observed distribution, and falls  back to the naive value predictor in case of previously unseen names.
  
\paragraph{Test generation}
This baseline uses Pynguin, a function-level test generator for Python~\cite{DBLP:conf/ssbse/LukasczykKF20}.
  For a fair, we run Pynguin on a single function at a time, which contains the code to execute only.
  Usually, test generators are applied to code with all dependencies fully resolved, where they are more effective than when being applied to a single function in isolation.
  The results of this baseline hence do not represent Pynguin's effectiveness in general, but in a usage scenario that matches that of \name{}.
  As many code snippets in the Stack Overflow dataset are not functions, we apply this baseline only to the open-source functions.    

\paragraph{Neural type prediction}
Because our value abstraction resembles types, this baseline builds on a recent stream of work on predicting type annotations in Python~\cite{Hellendoorn2018,fse2020,Allamanis2020,peng2022static,mir2022type4py}.
  We compare against the Type4Py model~\cite{mir2022type4py} because it outperforms earlier models, and because we can build on their publicly available, trained model.
  Type4Py predicts types for local variables defined in the code, as well as the parameters and return values of all functions. We concretize these types using the deterministic mapping in Table~\ref{tab:coarse-grained value abstraction}.
  For values where Type4Py does not predict any type, e.g., undefined variables, this baseline falls back to the random value predictor described above.

\subsection{RQ1: Accuracy of the Neural Model}

To measure the effectiveness of the model at predicting realistic runtime values, we compare the model's predictions against values observed during regular executions.
Specifically, we compare the predictions against a held-out subset of 5\% of the value-use events used for training the model.
These values serve as a ground truth because they are obtained by executing developer-written test suites, i.e., realistic executions performed without any guidance by \name{}.

For each of the two ways of representing concrete values (fine-grained abstraction and coarse-grained abstraction, Section~\ref{sec:value representation}), we train and evaluate a separate model.
To measure accuracy, we query the model for the most likely value in a given context and report how often this prediction exactly matches the ground truth (top-1).
In addition, we also report how often the ground truth value occurs in the top-3 and top-5 predictions by the model.

\begin{table}[]
  \caption{Accuracy of the neural model.}
  \label{tab:accuracy}
  \setlength{\tabcolsep}{8pt}
  \small
\begin{tabular}{@{}lrr|rr@{}}
\toprule
      & \multicolumn{4}{c}{Value abstraction}                                                                                 \\ \cmidrule{2-5} 
      & \multicolumn{2}{c}{Fine-grained}                          & \multicolumn{2}{c}{Coarse-grained}                        \\ \cmidrule{2-5} 
      & \multicolumn{1}{c}{CodeT5} & \multicolumn{1}{c}{CodeBERT} & \multicolumn{1}{c}{CodeT5} & \multicolumn{1}{c}{CodeBERT} \\ \cmidrule{2-5} 
Top-1 & 80.1\%                     &      79.5\%                      & 88.1\%                     &         87.3\%                     \\
Top-3 & 88.4\%                     &      94.5\%                      & 92.1\%                     &         96.5\%                    \\
Top-5 & 91.7\%                     &      96.8\%                      & 94.2\%                     &         98.2\%                    \\ \bottomrule
\end{tabular}
\end{table}

The results in Table~\ref{tab:accuracy} show that the neural models predict realistic values in the vast majority of cases.
The accuracy of the coarse-grained model ranges between 88.1\% and 94.2\% for CodeT5 and 87.3\% and 98.2\% for CodeBERT, depending on how many of the top-k predictions are considered.
Because the fine-grained models have a harder prediction task (i.e., more classes to choose from), their accuracy is slightly lower and ranges from 80.1\% to 91.7\% for CodeT5 and 79.5\% and 96.8\% for CodeBERT.
As we will see in RQ2, despite its lower accuracy, the fine-grained value abstraction is overall slightly more effective at guiding executions. Moreover, the accuracy per kind of value achieved by the models is close to the overall accuracy, e.g. the CodeT5 model with fine-grained value abstraction achieves 80.1\%, 84.1\%, and 76.0\% accuracy on name,  attribute, and  return value predictions, respectively.

To better understand the abilities and limitations of the neural model, we qualitatively study some of its mispredictions.
The most common cause is to predict a string when the ground truth value is \code{None}, and vice versa.
These are inherently difficult cases, because practically any value could be \code{None} in Python.
The second most common cause for mispredictions is that the model confuses lists and tuples, i.e., two kinds of values that often can be used interchangeably.
Inspecting specific examples of wrongly predicted values shows that the model is wrong mostly in situations where a human also cannot easily decide what the most realistic value is.
Specifically, we repeatedly observe two reasons.
First, the value is simply not known before loading it, e.g., in \code{if v is None:}, i.e., a branch depending on the value.
Second, the source code and its runtime behavior are inconsistent, e.g., when a docstring says that a function returns a dict, but actually returns a list.

\subsection{RQ2: Effectiveness at Covering Code}

\begin{figure}
  \centering
  \begin{subfigure}[b]{\linewidth}
      \centering
      \includegraphics[width=\textwidth]{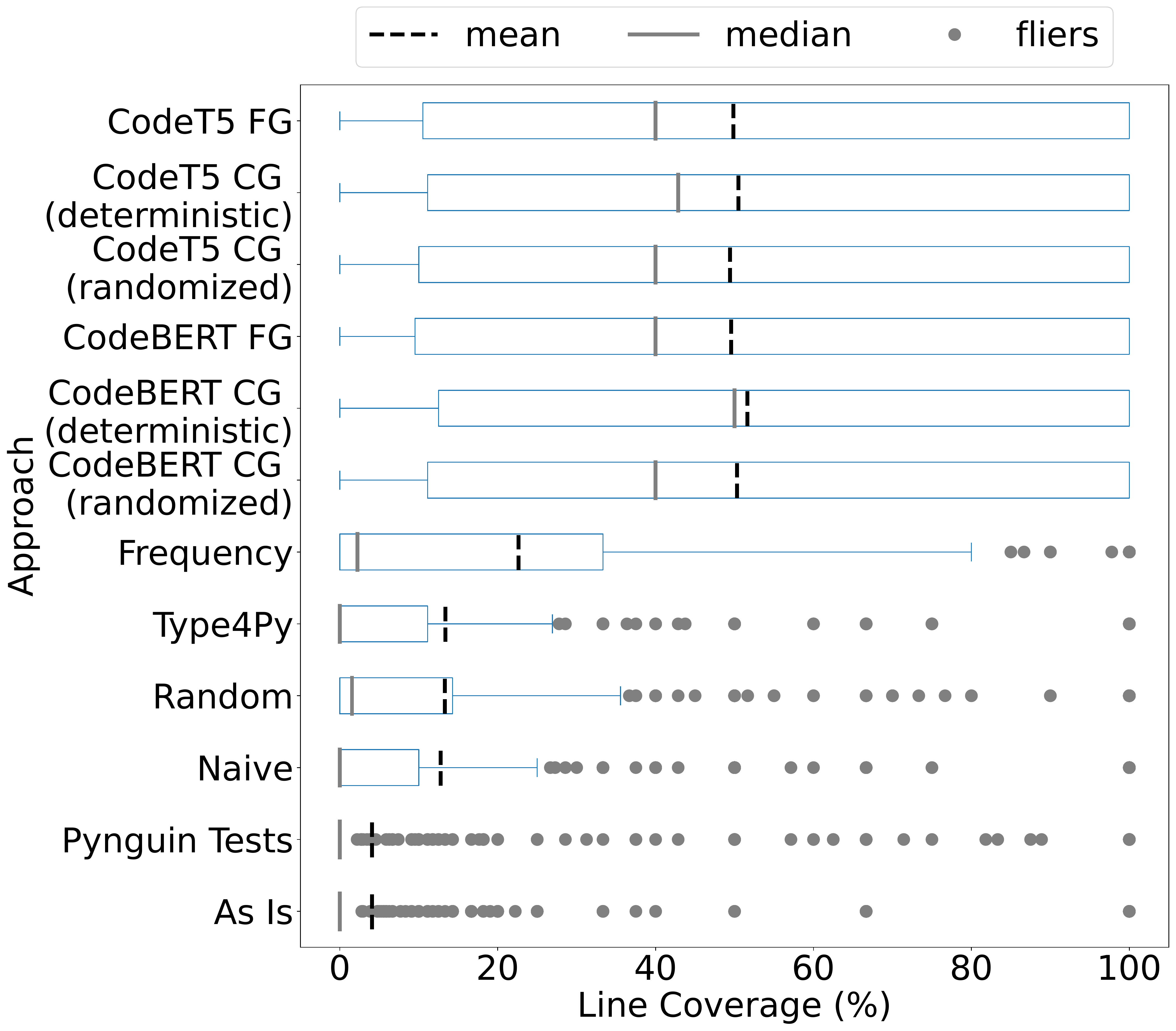}
     % \vspace{-1.7em}
      \caption{Open-source functions.}
      \label{fig:line_coverage_comparison_a}
  \end{subfigure}

  \vspace{1em}
  \begin{subfigure}[b]{\linewidth}
      \centering
      \includegraphics[width=\textwidth]{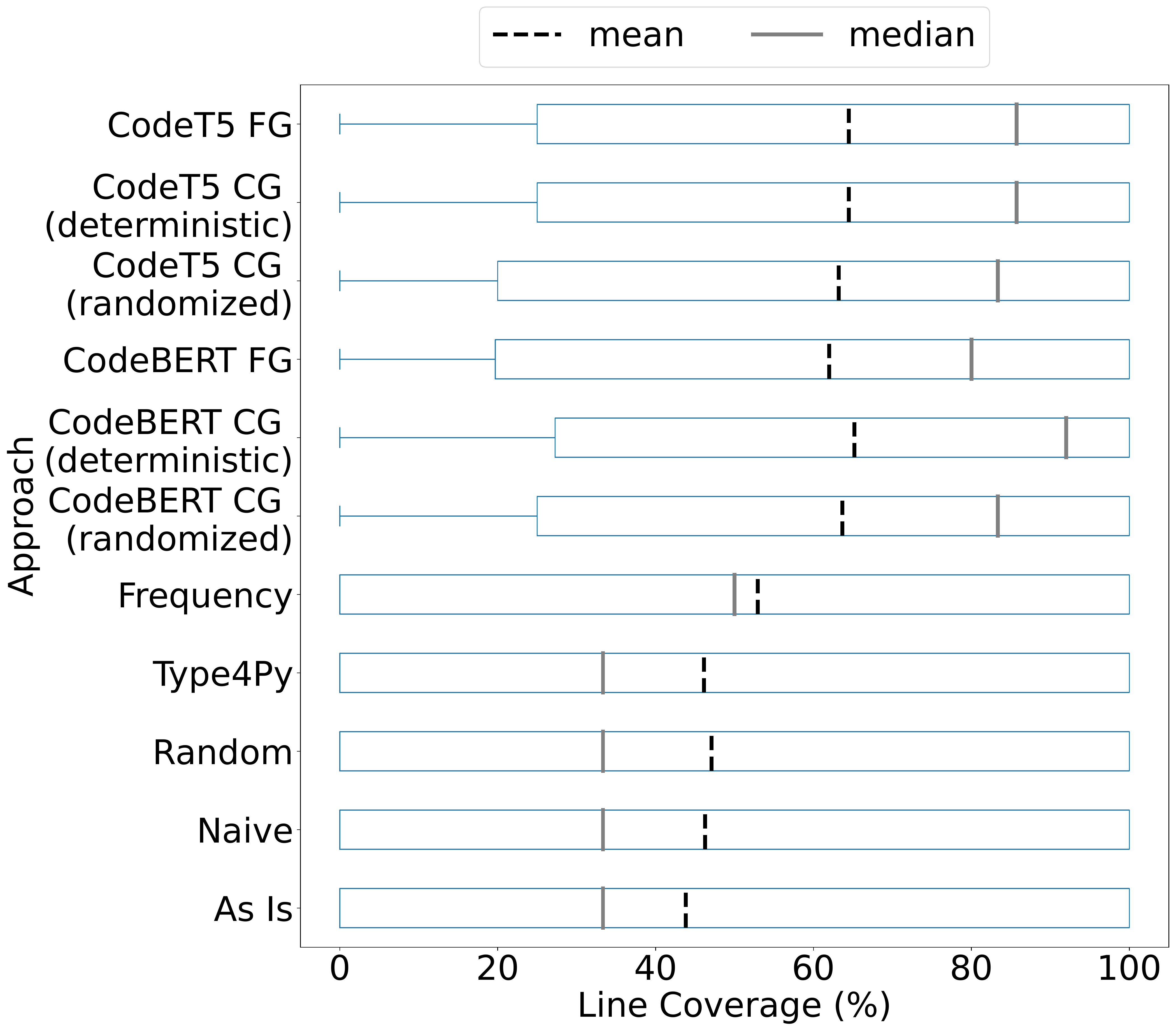}
      % \vspace{-1.7em}
      \caption{Stack Overflow snippets.}
      \label{fig:line_coverage_comparison_b}
  \end{subfigure}
         \caption{Line coverage on two datasets.}
     \label{fig:line_coverage_comparison}
     \vspace{-1.7em}
\end{figure}

To measure the effectiveness of \name{} at successfully executing code, we apply the approach to our two datasets and measure the percentage of all non-empty, non-comment lines of code that are successfully executed.
We call a line ``covered'' if the entire line executes without crashing.
We run all three variants of \name{}, as described in Sections~\ref{sec:value representation} and~\ref{sec:runtime engine}, using top-1 predictions: (i) fine-grained value abstraction, (ii) coarse-grained value abstraction with deterministic predictions, and (iii) coarse-grained value abstraction with randomized predictions.
We use the Wilcoxon signed-rank test ($p = 0.05$) to compare the significance of coverage differences between techniques. 

Figure~\ref{fig:line_coverage_comparison_a} shows the coverage achieved by \name{} and the baselines on the open-source functions.
On average, as-is execution and generated tests cover only \CoverageAsIsFcts{} of the lines (mean across all functions).
The naive, random, type (Type4Py), and frequency-based value predictors increase the mean coverage to 12.8\%, 13.3\%, 13.3\% and 22.6\%, respectively.
We attribute the negligible difference between the random predictor and the type predictor to the fact that the type predictor often falls back to the random predictor, because Type4Py does not predict any type for variables not defined in the given code (which is exactly the problem that \name{} addresses).
Finally, \name{} covers \CoverageFcts{} of the lines with CodeT5 and 51.6\% of the lines with CodeBERT, which is significantly higher than the other techniques.
Figure~\ref{fig:line_coverage_comparison_b} shows the coverage results on the Stack Overflow code snippets.
The coverage achieved by both the baselines and \name{} is higher, as many of these code snippets are meaningful without additional code context. On average, the open-source functions and the Stack Overflow code snippets contain 13 and seven missing values, respectively. For example, as-is execution and the random predictor cover \CoverageAsIsSO{} and 46.7\% of all lines, respectively.
Nevertheless, \name{} can significantly improve and achieves \CoverageSO{} coverage.

\begin{figure}
  \centering
  \begin{subfigure}[b]{\linewidth}
      \centering
      \includegraphics[width=\textwidth]{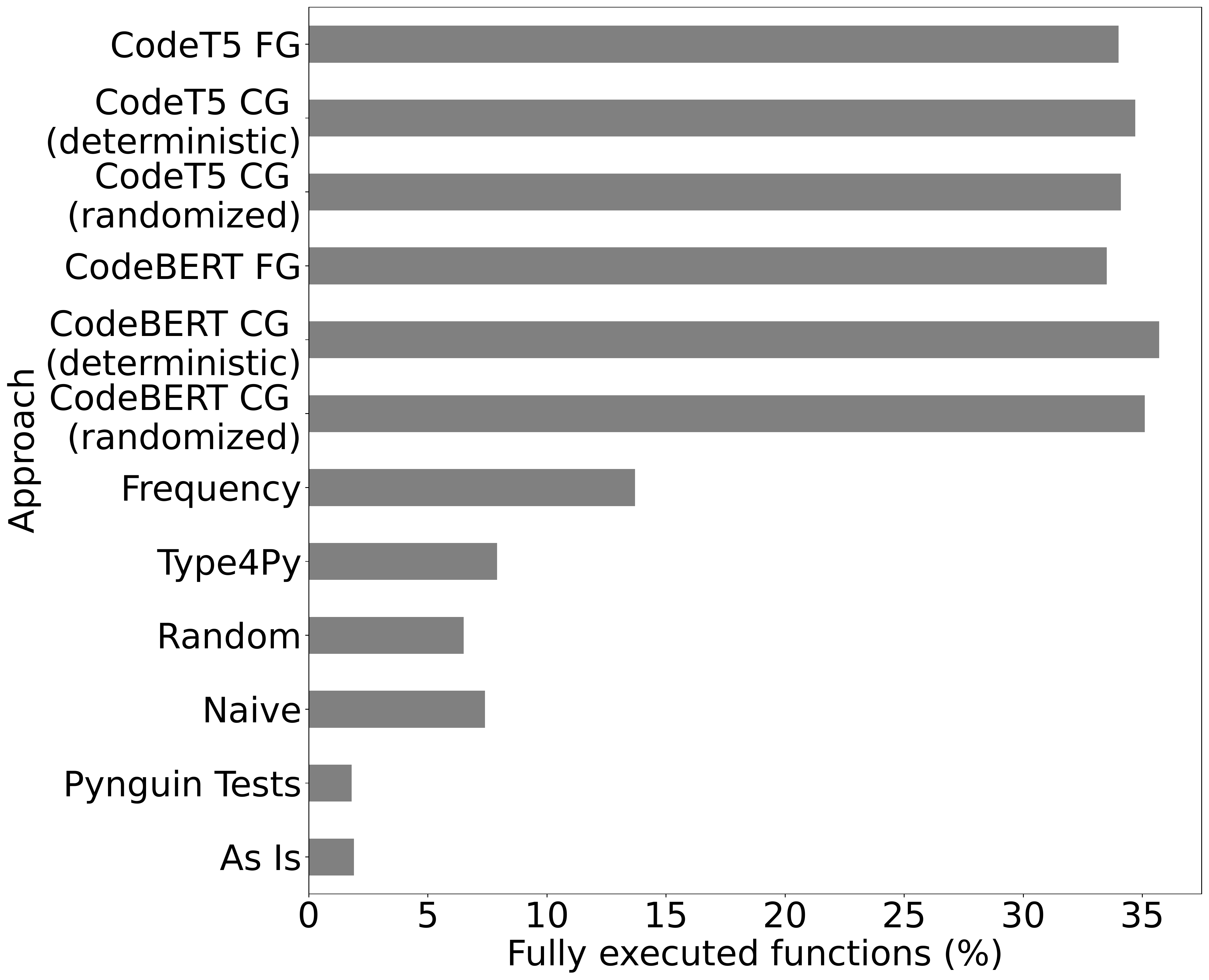}
     % \vspace{-1.7em}
      \caption{Open-source functions}
      \label{fig:full_coverage_comparison_a}
  \end{subfigure}

  \vspace{1em}
  \begin{subfigure}[b]{\linewidth}
      \centering
      \includegraphics[width=\textwidth]{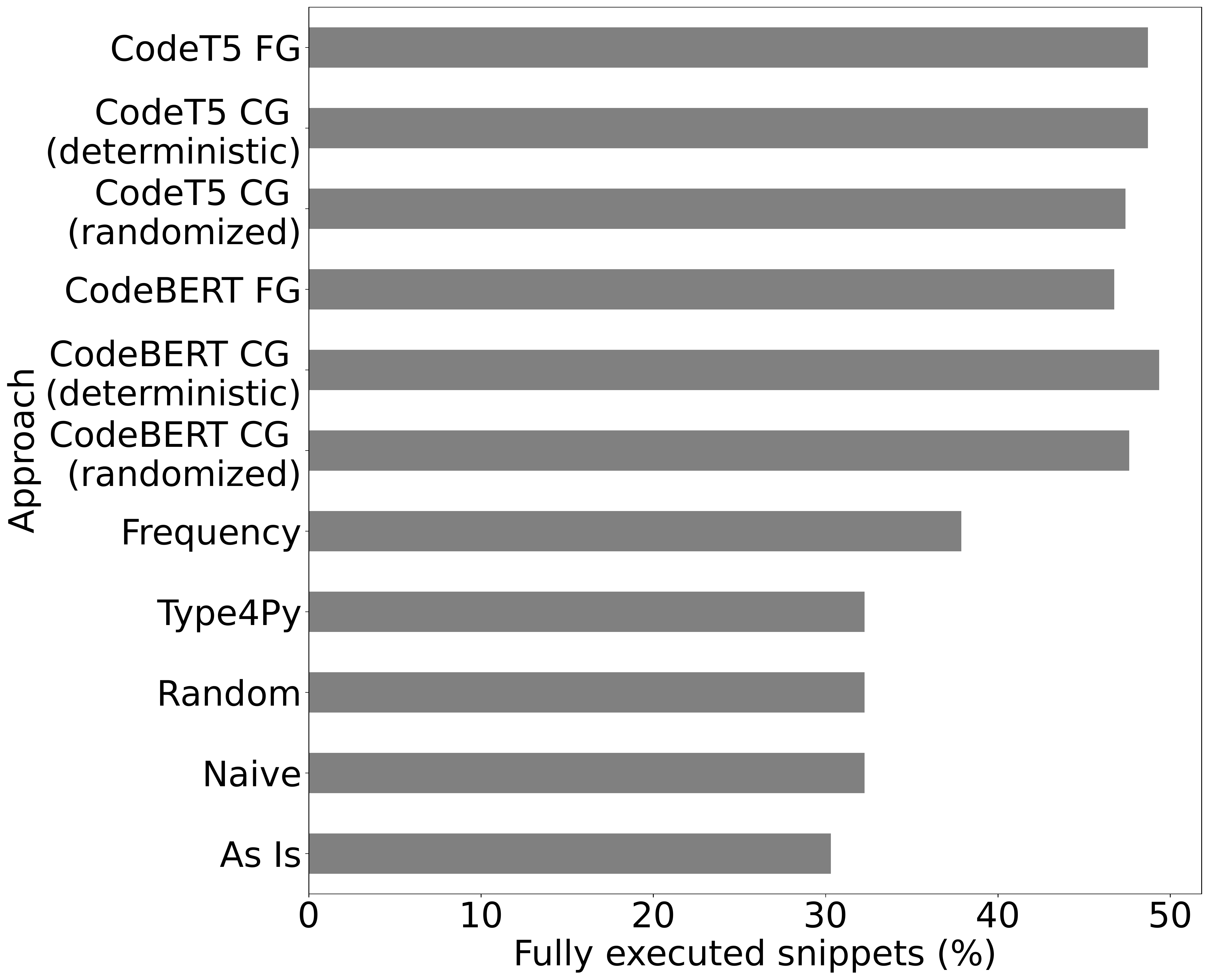}
     % \vspace{-1.7em}
      \caption{StackOverflow snippets}
      \label{fig:full_coverage_comparison_b}
  \end{subfigure}
     \caption{Percentage of code examples that are fully executed.}
     \label{fig:full_coverage_comparison}
     \vspace{-1.7em}
\end{figure}

\begin{figure}[t]
  \includegraphics[width=\linewidth]{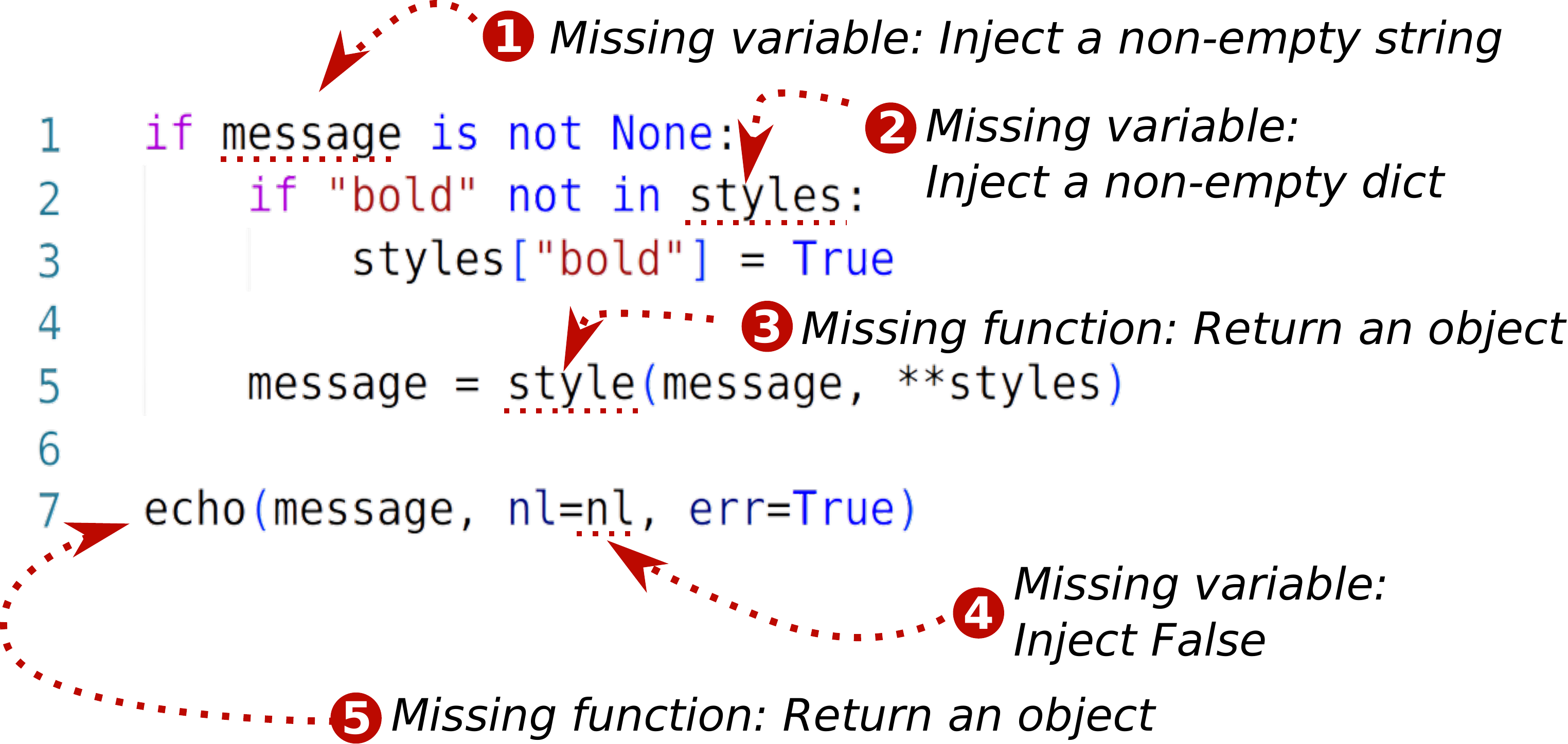}
  \caption{Open-source function code to execute and how \name{} guides the execution.}
  \label{fig:full_execution_example}
\end{figure}

Figure~\ref{fig:full_execution_example} shows the code of a function from the Black project, which is difficult to execute due to five missing pieces of information.
Running the code as-is crashes on the first line because the \code{message} variable is missing.
When running the code with the random value predictor, it will predict \code{message} to be an empty dictionary, \code{styles} to be a non-empty dictionary, and \code{style} to be an empty tuple.
These predictions cause a \code{TypeError: `tuple' object is not callable} exception at line~5.
In contrast, \name{} can fully execute the code by filling in the missing values with the realistic predictions shown on the figure.

% \begin{figure*}[t]
%   \includegraphics[width=.7\linewidth]{figs/partial_execution_example}
%   \caption{StackOverflow snippet to execute (left) and how \name{} guides the execution (right).}
%   \label{fig:partial_execution_example}
% \end{figure*}

As another metric of success, we measure how many of all code snippets we achieve 100\% line coverage.
Figure~\ref{fig:full_coverage_comparison} shows for how many of all considered code snippets the different approaches achieve this goal. 
Similar to the other coverage results, \name{} is more effective overall and clearly outperforms the baselines.
For example, for the open-source functions, \name{} fully executes the code of 35\% of all functions, whereas the best baseline (frequency value predictor) achieves this goal for only 13\% of the functions. 
These results reinforce that \name{} significantly improves over alternative approaches.

% \subsubsection{Examples}

% Figure~\ref{fig:partial_execution_example} shows a code snippet from StackOverflow.
% Similar to the previous example, running the code as-is crashes on the first line because \code{plt} is undefined.
% %
% \name{} instead predicts the values shown on the right side of Figure~\ref{fig:partial_execution_example}.
% With these predictions, the execution proceeds until line~5, where \name{} predicts a non-empty tuple for the missing variable \code{x}.
% Continuing with the execution, a \code{TupleError: tuple indices must be integers or slices, not tuple} exception is raised.
% The exception occurs because the comma syntax for slicing is not supported by any of the built-in sequence types in Python.%, but only by numpy arrays.
% Given that the \code{x} parameter of the \code{matplotlib.pyplot.hist} function is expected to be an array or a sequence, \name{}'s prediction for the variable \code{x} is realistic, and the exception might indeed occur in practice.

\subsection{RQ3: Efficiency at Guiding Executions}

We evaluate the efficiency of learning-guided execution by measuring (i) the time required to instrument the code and (ii) the average time to execute a line of code.
The instrumentation takes 38.9 seconds for the 1,000 functions in Table~\ref{tab:random_functions_projects} and 20.6 seconds for the 462 Stack Overflow snippets.
% 9653+3580=13233 LoC  -->  4.5 seconds per 1,000 LoC
That is, instrumenting 1,000 lines of code takes 4.5 seconds, on average.
% Parallelizing the instrumentation could easily reduce this time further.
%
Table~\ref{tab:efficiency-results} summarizes the time for executing code within \name{} and with the baselines.
As expected, as-is execution is the most efficient, as it runs code in the bare Python interpreter.
The three baselines that inject values without querying the neural model, i.e., naive, frequency-based, and random value prediction, moderately slow down the execution, for example by a factor of roughly 2x for the functions dataset.
Because \name{} queries the neural model for each missing value, it takes the longest to execute the code, and causes a slowdown of one to two orders of magnitude.
Taking the results from RQ2 and RQ3 together, an increased execution time is the cost to pay for being able to execute many lines of code at all.
We consider \name{}'s efficiency to be acceptable for many dynamic analysis applications, as dynamic analyses often impose similar overheads~\cite{fse2022-DynaPyt,asplos2019,Sen2013}.

\begin{table}[]
\caption{\label{tab:efficiency-results} Average execution time (ms) per LoC.}
\centering
\setlength{\tabcolsep}{7pt}
\small
\begin{tabular}{@{}lrr@{}}
\toprule
Approach            & \multicolumn{2}{c}{Dataset}                                                       \\
\cmidrule{2-3}
& Functions & Stack Overflow \\
\midrule
CodeT5 FG                 & 178.69                        & 47.29 \\
CodeT5 CG (deterministic) & 185.08                                 & 46.23
 \\
CodeT5 CG (randomized)    & 167.48                                & 46.31                                       \\

CodeBERT FG               & 464.83                       &   133.76   \\

CodeBERT CG (deterministic) &  479.89                    &   126.47  \\

CodeBERT CG (randomized) &   438.64                     &     127.20    \\

Random                                   & 3.94                                & 5.93
 \\
Frequency                                & 3.61                                & 5.73
 \\
Naive                                    & 3.62                                & 5.42
 \\
As Is                                    & 1.50                                 & 5.19                                       \\
\bottomrule 
\end{tabular}
\end{table}

\subsection{RQ4: Using \name{} to Find Semantics-Changing Commits}

To illustrate an application of our approach, we apply \name{} to find semantics-changing commits.
Finding such commits is useful to understand whether a commit that changes a function's implementation also changes its behavior.
For example, if a commit described as a refactoring is found to actually be semantics-changing, then this commit could be prioritized for manual inspection.
However, executing the code involved in a commit usually is possible only if a test suite exists that covers the modified code.

We use \name{} to execute the code of a function before and after a commit, and then report the commit as semantics-changing if the values returned by the two functions differ.
As a dataset, we extract from the 1,000 most recent commits of each of the projects in Table~\ref{tab:random_functions_projects} those commits that change a single function from $f_{old}$ to $f_{new}$.
For each such pair, we generate a file that contains $f_{old}$, $f_{new}$, and code that invokes both functions and then compares their return values.
The consistency mechanism described in Section~\ref{sec:runtime engine} ensures that \name{} predicts the same values for those parts of the functions that have not changed.
% For example, if $f_{old}$ is missing a value for a variable \code{name} and the approach injects a non-empty string, then the same value is also injected into $f_{new}$ when it refers to the variable \code{name}.
%
We consider two return values to be different if (i) they have different types, (ii) both have a primitive type, e.g., \code{int} or \code{str}, but a different value, or (iii) if both values are collections, e.g., a list or a set, but differ in size.
In contrast, we do not report a difference if the two values are different objects, as accurately comparing complex objects is difficult.

Comparing executions may lead to three possible outcomes:
\begin{enumerate}
  \item \emph{Exceptional}. At least one of the functions raises an exception.
  We do not draw any conclusion in this case because the exception may be due to an unrealistic value injected by \name{}.
  \item \emph{Same behavior}. Both functions return the same value, i.e., that the approach has not found any change in the semantics.
  \item \emph{Semantics-changing}. The two functions return different values, i.e., the approach has detected a semantics-changing commit.
\end{enumerate}

\begin{table}
  \caption{Results from finding semantics-changing commits.}
  \label{tab:commit analysis}
  \setlength{\tabcolsep}{8pt}
  \small
  \begin{tabular}{@{}lrrrr@{}}
    \toprule
Project & \multicolumn{4}{c}{Commits} \\
\cmidrule{2-5}
& Total & Exceptional & Same & Semantics- \\
&  &  & behavior & changing \\
    \midrule
Black & 68 & 41 & 27 & 0 \\
Flask & 114 & 78 & 36 & 0 \\
Pandas & 611 & 403 & 207 & 1 \\
Scrapy & 522 & 292 & 220 & 10 \\
TensorFlow & 320 & 241 & 77 & 2 \\
    \midrule 
Total & 1,635 & 1,055 & 567 & 13 \\
    \bottomrule
  \end{tabular}
\end{table}

\begin{figure}
\begin{lstlisting}[numbers=left,xleftmargin=12pt]
def _retry(self, req, reason, spider):
  retries = req.meta.get('retry_times', 0) + 1

  (@\HL@)if 'max_retry_times' in req.meta:(@\HLoff@)
    (@\HL@)self.max_retry_times = req.meta['max_retry_times'](@\HLoff@)

  stats = spider.crawler.stats
  if retries <= self.max_retry_times:
    logger.debug("Retrying ...")
    # (more code)
    return retryreq
  else:
    stats.inc_value('retry/max_reached')
    logger.debug("Gave up ...")
\end{lstlisting}
  \caption{Commit found to be semantics-changing (the highlighted code was newly added).}
  \label{fig:example commit}
  \vspace{-1.7em}
\end{figure}

Table~\ref{tab:commit analysis} shows the results.
The approach finds 567 commits that preserve the behavior and 13 commits that are semantics-changing.
Figure~\ref{fig:example commit} shows an example.
The commit adds lines~4 and~5, which may update the value of \code{max\_retry\_times}.
This value influences the branching decision at line~8, which in turns determines whether the function returns in object or \code{None}.
Usually, executing the old and new code in isolation is impossible, as it requires complex input objects and refers to imported functions, but with \name{}, we can successfully execute the code.
Moreover, the execution of the old code takes the branch at line~8 and hence returns an object, whereas the execution of the new code enters the \code{else} branch and hence returns \code{None}, i.e., the commit is semantics-changing.

% Example candidates:
% https://github.com/scrapy/scrapy/commit/effaab867e45d441b5ca99ea6b24fd423e256a33
% https://github.com/scrapy/scrapy/commit/810658bcc5b1897d57c0882a0f7ab4a0d264a778

Being based on dynamic analysis, the approach underapproximates the semantics-changing commits.
That is, not finding any difference does not guarantee that a commit is semantics-preserving, as another execution might still expose a behavioral difference.

\section{Discussion}

\paragraph{Threats to Validity}
As a ground truth for judging whether a value is realistic (RQ1), we use values observed during regular executions.
However, sometimes more than one kind of value may realistically appear at a given code location, e.g., \code{None} vs.\ some other value.
As a result, our accuracy metric underestimates the ability of the model to predict realistic values.
We implement \name{} for Python and hence cannot conclude how well our ideas would work in another language.
We believe that the approach could easily be adapted to other languages though.
In particular, a statically typed language, such as Java or C++, even provides additional information about the values to inject, which would be used to constrain the model's search space.
As usual, all conclusions are limited to the datasets we use.
To mitigate this threat, we select datasets that cover different usage scenarios and different applications domains.

\paragraph{Limitations}
Because \name{} offers a form of underconstrained execution, it cannot guarantee that an execution may occur in practice.
Further improvements of the neural model could address this limitation.
In addition, future work could reduce the risk to diverge from a realistic execution by combining \name{} with executions of existing tests or with constraint-based reasoning.
Another limitation is to inject only a relatively small set of concrete values (Tables~\ref{tab:value abstraction} and ~\ref{tab:coarse-grained value abstraction}), which limits our ability to inject realistic values.
For example, given a program that usually reasons about strings that contain email addresses, \name{} can at best inject a string \code{"a"}, but would never inject a valid email address.
We plan to explore continuous value representations and a richer concretization mechanism in the future.

\section{Related Work}

\paragraph{Neural software analysis}
Analyzing software using neural networks has gained a lot of traction over the past few years~\cite{NeuralSoftwareAnalysis}.
One key question is how to represent the code in a way that enables effective neural reasoning, with answers including techniques for representing individual tokens~\cite{Karampatsis2020,icse2021,Chen2022}, token sequence-based models~\cite{Gupta2017,Hellendoorn2018,Chen2019}, AST-based models~\cite{Mou2016,Alon2019,Zhang2019}, and models that operate on graph representations of code~\cite{Allamanis2018b,Dinella2020,Wei2020}.
Neural models of code are shown to be effective for various tasks, e.g.,
bug detection~\cite{oopsla2018-DeepBugs,Li2018a,Hellendoorn2020,Vasic2019},
code completion~\cite{DBLP:journals/corr/abs-2004-05249,DBLP:conf/icse/KimZT021,Alon2019a},
code summarization~\cite{Allamanis2015,Allamanis2016,Alon2019a,Zhang2020,Ahmad2020,Liu2021},
program repair~\cite{Gupta2017,Dinella2020,Chen2019,Tarlow2019,Lutellier2020,Li2020a,Yasunaga2021,Ye2022a},
type prediction~\cite{Hellendoorn2018,icse2019,fse2020,Wei2020,Allamanis2020},
and code search~\cite{Sachdev2018,Gu2018,DBLP:conf/icse/SunFCTHZ22}.
More recently, researchers have started to explore the predictive power of large-scale, pre-trained models~\cite{Feng2020,Guo2021,DBLP:conf/emnlp/0034WJH21,Chen2021,Xu2022}, which reduce the per-task training effort, either via fine-tuning~\cite{Chakraborty2021a,Pei2021} or few-shot learning~\cite{codexStudy2022,Jain2022,Poesia2022}.
\name{} differs by making predictions about runtime values, and by using these predictions to manipulate an ongoing execution.

\paragraph{Neural models of executions}
A few neural software analyses focus on executions.
Some models predict the behavior of an entire program~\cite{Zaremba2014,DBLP:conf/nips/BieberSLT20}, but are limited to small, synthetically created code snippets that do not suffer from any missing values.
Other work passes runtime traces into models~\cite{Wang2020,Pei2021}, predicts whether a code snippet will raise an exception~\cite
{Bieber2022}, uses runtime information as feedback during training ~\cite{Ye2022a}, and ask a model to produce a trace of intermediate computation steps~\cite{Nye2021}.
Nalin spots inconsistencies between values and the names that refer to them~\cite{icse2022-Nalin}.
Their work exploits the predictability of many runtime values to identify outliers, whereas we exploit this property to fill in otherwise missing values.
Compared to all the above work, \name{} contributes by letting a model predict runtime values to be used when the program would crash otherwise.

\paragraph{Test generation}
\name{} and test generators share the goal of making incomplete code run.
Popular approaches include Quick-\\Check~\cite{claessen2000quickcheck},
JCrasher~\cite{Csallner2004}, 
symbolic execution~\cite{Cadar2008,Sen2005,Godefroid2005},
Randoop~\cite{Pacheco2007},
EvoSuite~\cite{Fraser2011a},
and AFL~\cite{afl2013,Boehme2019}.
% Fuzzing can also benefit from inferred grammars~\cite{DBLP:journals/tse/SoremekunPHGZ22}.
Finally, there are test generators for specific kinds of code, e.g., higher-order functions~\cite{oopsla2018-LambdaTester}, code with asynchronous callbacks~\cite{icse2022-Nessie}, or parsers~\cite{DBLP:conf/pldi/MathisGMKHZ19,DBLP:conf/issta/MathisGZ20}.
A key difference to our work is that \name{} injects values at arbitrary points in the execution on-demand, instead of injecting values at a well-defined interface, such as a function entry point.
% In the context of testing, our work could also be described as on-demand mocking, as \name{} provides missing values in a way similar to mocking.

\paragraph{Underconstrained execution}
Usually, every value the program operates on is known, or otherwise the program gets stuck and is aborted.
In contrast, prior work proposes several ways of executing a program without having to fully define all values the program operates on, i.e., forms of underconstrained execution:
micro-execution~\cite{Godefroid2014}, which allows for executing arbitrary x86 code by injecting values into memory on demand;
% Trex~\cite{DBLP:journals/corr/abs-2012-08680} uses traces obtained via micro-execution to train a model that reasons about the runtime semantics of code to identify similar functions.
%
underconstrained symbolic execution~\cite{Ramos2015}, which applies symbolic execution to individual functions in isolation, instead of entire programs; and
forced execution~\cite{Kim2017}, which artificially forces the program to take not yet explored branches.
Our work fundamentally differs from the above by predicting realistic instead of arbitrary values.

\paragraph{Other related work}
Some static analyses support code with missing or incomplete type information~\cite{Dagenais2008a,DBLP:journals/pacmpl/MeloRAP18}.
In contrast, our work enables dynamic, not static, analysis.
Other work heuristically resolves missing dependencies~\cite{Phan2018,DBLP:conf/icse/WangLZ21}.
\name{} addresses the complementary problem of predicting missing values.
A study~\cite{DBLP:journals/corr/abs-1907-04908} finds most Python code on  Stack Overflow to be non-executable, which \name{} addresses at least partially.
%
% "DockerizeMe: Automatic Inference of Environment Dependencies for Python Code Snippets" has more references for code snippets being often not executable
%
As our work enables executing otherwise non-executable code, it will enable dynamic analyses for Python~\cite{fse2022-DynaPyt}.

%\todo{cite CodeExecutor by Microsoft}

\section{Conclusions}

Motivated by the recurring need to execute incomplete code, such as code snippets posted on the web or code deep inside a complex project, this paper introduces \name{}.
Our approach prevents exceptions due to missing values, and instead, queries a neural model to predict likely values to use instead.
The result is a novel technique for executing code in an underconstrained way, i.e., without a guarantee that the execution reflects real-world behavior, but with a much higher chance of executing without crashing.
% Because the neural model underlying \name{} predicts realistic values with a high accuracy (\ModelAccuracyMin{} to \ModelAccuracyMax{}), the approach effectively guides the execution toward covering \CoverageFcts{} of the code in open-source functions and \CoverageSO{} of code posted on Stack Overflow, both of which clearly improve upon existing baselines.
We envision our approach to provide a foundation for various dynamic analysis applications, because it facilitates the execution of otherwise impossible or difficult to run code.

\section{Data Availability}

Artifact: \url{https://zenodo.org/record/8270900}\\
Newest version: \url{https://github.com/michaelpradel/LExecutor}

% Our implementation is available as open-source at:
% \begin{center}
%   \url{https://github.com/michaelpradel/LExecutor}
% \end{center}

\begin{acks}
  % \vspace{.3em}
  % \footnotesize
  % \begin{spacing}{0.8}
This work was supported by the European Research Council (ERC, grant
agreement 851895), by the German Research Foundation within the
ConcSys and DeMoCo projects, and by the National Council for Scientific and Technological Development (CNPq)/Brazil (Process 162049/2021-8).
  % \end{spacing}
\end{acks}

\bibliographystyle{ACM-Reference-Format}
\bibliography{referencesMP,referencesMore}

\end{document}